\documentclass[narrowdisplay]{elsart}
\usepackage[english]{babel}
\usepackage{graphicx}
\usepackage{overcite}
\usepackage{epstopdf}

\begin{document}

\begin{frontmatter}
\enlargethispage{1\baselineskip}
\title{Optical characterization of PZT thin films for waveguide applications}
\author{J. Cardin\corauthref{cor}},
\ead{julien.cardin@physique.univ-nantes.fr}
\author{D. Leduc\corauthref{cor}},
\ead{dominique.leduc@physique.univ-nantes.fr}
\corauth[cor]{Corresponding authors. Tel.:+33-2-5112-5533; Fax: +33-2-4014-0987}
\author{T. Schneider, C. Lupi, D. Averty and H. W. Gundel}
\address{\it Institut de Recherche en Electrotechnique et Electronique de Nantes-Atlantique (IREENA), Facult\'e des Sciences et des Techniques, Universit\'e de Nantes, 2 rue de la Houssini\`ere, 44322 Nantes Cedex 3, France}
\begin{abstract}
In order to develop an electro-optic waveguide, $Pb(ZrTi)O_3$ ceramic ferroelectric thin films were elaborated by a modified sol-gel process on glass substrate. In the aim to study the optical properties of the PZT films, an accurate refractive index and thickness measurement apparatus was set-up, which is called M-lines device. An evaluation of experimental uncertainty and calculation of the precision of the refractive index and thickness were developed on PZT layers. Two different processes of PZT elaboration were made and studied with this apparatus. The reproducibility of one fabrication process was tested and results are presented in this paper.
\end{abstract}
\end{frontmatter}

\section{Introduction}
In Integrated Optics, various applications are based on the utilisation of thin film planar waveguides. The approach to use PZT transparent ceramic films for the realization of such waveguide structures will be reported in the present paper. The PZT thin films were elaborated by a Chemical Solution Deposition (C.S.D.) technique and were spin-coated on glass substrates. The main characteristic parameters for the realization of a planar waveguide are the refractive indexes and the thicknesses of the different layers constituting the waveguide$^1$. In order to accurately measure these parameters, a prism-film coupler device was used (so-called M-lines technique) at the 632.8 nm He-Ne laser wavelength allowing to determine the phase matching angles (coupling modes). Numerical resolution of the planar waveguide equation allows to relate these angles to the mode order, to the refractive index, and the film thickness. Evaluation of different uncertainty of our experimental set-up and a differential calculus allows to quantify the accuracy of the refractive index and the film thickness.
\section{Film deposition and characterization}

\begin{figure}[htbp]
\centerline{
\includegraphics[width=7cm]{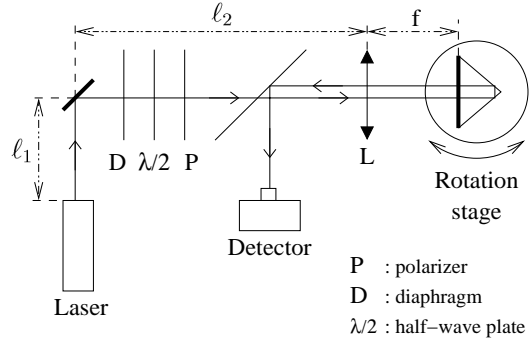}
}
\caption{M-lines experimental set-up: $\ell_1=13\ cm$, $\ell_2=59\ cm$, and $f=20\ cm$.}
\label{fig1}
\end{figure}

Pb$_{1.4}$Zr$_{0.36}$Ti$_{0.64}$O$_3$ thin films were elaborated by a (modified Sol-gel process) Chemical Solution Deposition (C.S.D.) technique and spin-coated on glass substrates (Corning 1737f). The precursor solution consisted of lead acetate dissolved in acetic acid, zirconium and titanium n-propoxide; ethylene glycol was added in order to prevent from crack formation during the annealing process. The deposited films were dried on a hot plate and a Rapid Thermal Annealing (RTA) procedure at 620$^\circ$C resulted in the formation of a polycrystalline perovskite phase$^{2,3}$. The substrate has been selected because of its thermal expansion coefficient close to this of bulk Pb(ZrTi)O$_3$. This minimizes thermal strains in the film during the heating process, reduces the number of cracks and consequently increases the optical quality of the thin films. The structure of the films was characterized by SEM and X-ray diffraction and their optical properties were investigated with M-lines spectroscopy$^4$.
We will focus on this optical characterization because the refractive index is one of the major parameters for the design of waveguides. 
A classical M-lines device was set-up in the configuration shown in figure \ref{fig1}.
The light of a He-Ne laser is focused to the base of a prism, which is pressed against the film. The prism and the film are mounted on a rotating stage allowing to vary the angle of incidence. The incident light is totally reflected by the interface prism-film for all angles except synchronous angles
where it is coupled into the film. For TE modes, these synchronous angles are given by the dispersion equation of planar waveguides~:
\begin{equation}
\begin{array}{ll}
m\pi= & dk_0\sqrt{n^2-N_m^2}
       -\mbox{arctg}\left(\sqrt{\frac{N_m^2-n_a^2}
                          {n^2-N_m^2}}\ \right)\\[4mm]
      &   -\mbox{arctg}\left(\sqrt{\frac{N_m^2-n_s^2}
                          {n^2-N_m^2}}\ \right)
\end{array}
\label{eq1}
\end{equation}
where $m$ is the order of the guided mode, $d$ is the thickness of the film,
$n$, $n_a$, $n_s$ are respectively the film, the air and the substrate refractive index and
$N_m$ is the effective index of the guided mode. $N_m$ depends on the
synchronous angles $\phi_m$
\begin{equation}
N_m=n_p\sin\left[A_p+\arcsin\left(\frac{\sin\phi_{m}}{n_p}\right)\right]
\label{eq2}
\end{equation}
where $A_p$ is the angle of the prism and $n_p$ is its refractive index. 
In the preceding equations, the two parameters $n$ and $d$ are unknown.
Hence, it is necessary to measure at least two modes in order to determine them$^5$.

The precision of the method is determined by the precision of the angle measurement.
The rotation is done by a step by step motor having a step size of 0.001$^\circ$, which can be considered 
as the maximum precision. In the real experiment, other uncertainties lower this precision, the most important of which is the uncertainty on the position of the zero angle. Experimentally, this position was defined by superimposing the backward and forward beams on the diaphragm (see figure \ref{fig1}). In order to evaluate numerically this uncertainty, we suppose that the beams are superimposed if their centers are closer than one radius. With the focal length of the lens L of 20~cm and a beam waist of the laser of 800 $\mu m$, an uncertainty of the zero angle position of 0.05$^\circ$ is found.

The influence of this error on the values of $n$ and $d$ is calculated whith the help of a set of two dispersion equations (\ref{eq1}) for two modes $m_1$ and $m_2$. By differentiating these equations, we obtain the uncertainty on $n$ and $d$ as a function of the uncertainties of the other parameters~: $ \Delta n=G_n(\Delta n_a,\Delta n_s,\Delta N_{m1},\Delta N_{m2})$ and $ \Delta d=G_d(\Delta n_a,\Delta n_s,\Delta N_{m1},\Delta N_{m2})$, where $ \Delta N_{mi}$ depends on $\Delta A_p$, $\Delta n_p$ and $\Delta \phi_{mi}$.

\begin{figure}[htbp]
\centerline{
\includegraphics[width=7cm]{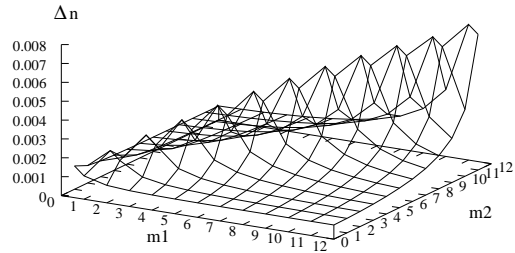}
}
\caption{Index uncertainty as a function of the two mode orders.}
\label{fig2}
\end{figure}

The figure \ref{fig2} shows the uncertainty of refractive index as a function of the mode orders for a 2.08 $\mu m$ thick film having an index of 2.427 (the M-lines spectrum is represented in figure 3) for angle uncertainties of 0.05$^\circ$.
The uncertainty of the refractive index increases with the mode orders, however remains smaller than $\pm 8.10^{-3}$.
\section{Mode indexing}
\begin{figure}[htbp]
\centerline{
\includegraphics[width=7cm]{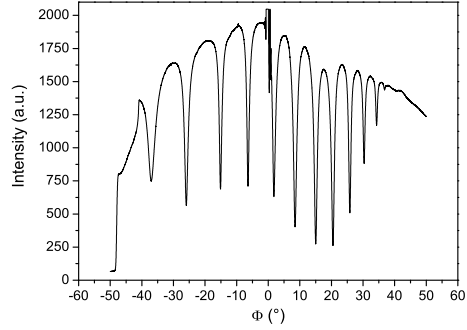}
}
\caption{M-lines spectrum of a 2.08 $\mu m$ thick pyrochlore film.}
\label{fig3}
\end{figure}

\begin{figure}[htbp]
\centerline{
\includegraphics[width=7cm]{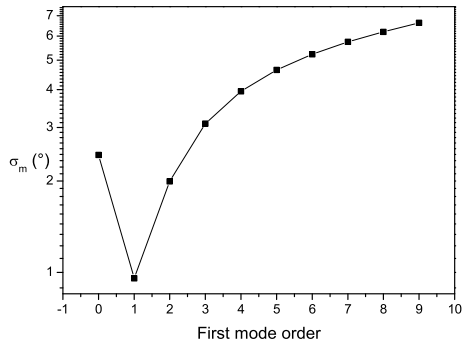}
}
\caption{The $\sigma_m$ curve as a function of the first mode order.}
\label{fig4}
\end{figure}

Usually, mode indexing is not a very difficult task.
Under certain circumstances, however, especially when developing
a film processing method, great care must be taken since the validity of the dispersion equation is restricted to homogeneous, isotropes and losless guides. Films that deviate from these ideal conditions could give erroneous results. The criterion usually used for indexing states that the first appearing mode is the zero order mode. In the case of real films, however it is sometimes very difficult to excite this fundamental mode because of the large angle of incidence. The peak corresponding to the zero order mode is then missing in the M-lines spectrum. This is illustrated in the figure \ref{fig3} which shows the M-lines spectrum of a Pb$_{1.4}$Zr$_{0.36}$Ti$_{0.64}$O$_3$ thin film cristallized in the pyrochlore phase. In our setup, the fundamental mode appears at high angles. A tiny peak around 37$^\circ$ can be 
seen which is hardly measurable and appeared to be insensitive to the pressure applied to the film. Thus, it is not certain that this peak is the fundamental mode. In order to overcome the mode order indexation difficulty, we suggest the following procedure. For a spectrum consisting of M modes, we suppose that the first mode is equal to zero, the second to one and so on. Using the equation \ref{eq1}, we calculate the $\rm {C}_M^2$ couples $(n,d)$ of solutions corresponding to the different combinations of two modes. With this couples (n,d) of solutions, using the same equation, M synchronous angles $\phi^{\rm calc}$ are obtained. Then we build the $\sigma_m$ fonction (Equation \ref{eq3}) which is the square root of sums of each mode and of each combination of the quadratic difference between these $\phi^{\rm calc}$ and the measured synchronous angles $\phi^{\rm meas}$:
\begin{equation}
\sigma_m=\sqrt{\sum_{i=1}^{\rm C^2_M}\ \sum_{j=1}^{M}\frac{\left|\phi^{\rm calc}_{ij}-\phi^{\rm meas}_{ij}\right|^2}{M\ \rm C^2_M}}
\label{eq3}
\end{equation}
This calculus is repeated several times while incrementing each mode by one unit, i.e. in the second step the first mode is equal to one ($m=1$), the second is equal to two and so on. Applying this procedure in order to analyse the spectrum of figure \ref{fig2}, we find a $\sigma_m$ as a function of the first mode order as shown in figure \ref{fig3}. Numerical simulations showed that the smallest $\sigma_m$ results in a correct indexing for angles measurement errors smaller than 0.5$^\circ$. On the figure \ref{fig4}, the minimum appears for $m=1$, thus, the first peak of the spectrum (around $37^{\circ}$) is not the fundamental mode.

Once the indexing is finished, an optimisation method is necessary in order to find the actual refractive index and the thickness of the film because of the dispersion of values resulting from the different mode combinations.
The simplest method is to calculate mean values, however, for a number greater than four, we prefer to use a simplex algorithm which allows to reduce systematic errors of the origine of the angles$^6$. In the case of the studied film, this optimization gives the following results~:
$n=2.427\pm2.10^{-3}$ and $d=2.08\pm2.10^{-2}~\mu m$.
  
\section{Study of ferroelectric films}

Convenable waveguides are realised from homogeneous films of the order of $1\mu m$ thickness. 
The difficulty which arises for this type of layers is that cracks may appear during the cristallisation process. In order to limit these cracks, we compared two cooling methods after the heat treatment, one outside the oven and one inside the oven. No apparent differences were visible from a macroscopic point of view.

\begin{figure}[htbp]
\centerline{
\includegraphics[width=7cm]{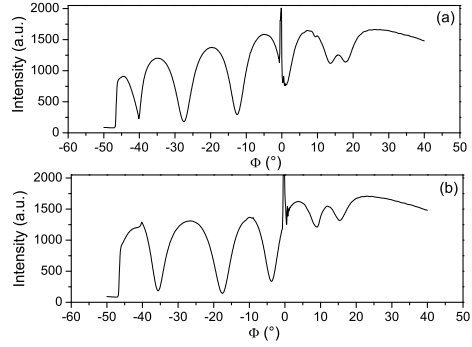}
}
\caption{M-lines spectrum on two PZT sample elaborated with the both method: (a)Cooling outside the oven (b)Cooling inside the oven.}
\label{fig5}
\end{figure}

\begin{figure}[htbp]
\centerline{
\includegraphics[width=7cm]{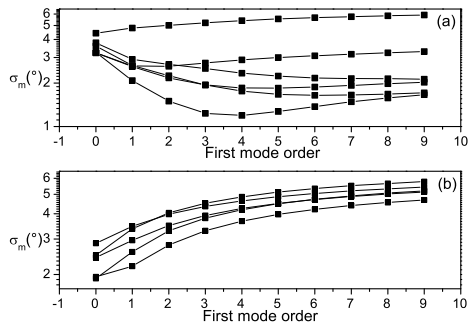}
}
\caption{$\sigma_m$ curves as a function of the first mode order for: (a)Cooling outside the oven, (b)Cooling inside the oven.}
\label{fig6}
\end{figure}

The M-lines spectra obtained for the two methods are represented in the figure \ref{fig5}. 
Despite of only a modification in the heat treatment the M-lines spectra show considerable differences in the position of the individual peaks. A series of six identical samples was prepared for each cooling procedure, the M-lines spectra were acquired, and the above described indexing method was applied. The square root of sums $\sigma_m$ for cooling outside and inside the oven is given in figure \ref{fig6}a and \ref{fig6}b, respectively, showing a qualitative difference of the two series.

For the films having undergone a cooling outside the oven, it is either impossible to find a minimum, or the indexing obtained leads to absurd values of $n$ and $d$. Apparently, the thermal shock imposed to the samples while retrieving them from the oven, results in a modification of the films, which is macroscopically invisible. However, it can be detected by M-lines spectroscopy. While no direct evidence on the strucural properties of the films may be derived from this measurements, the results show a nonconformity between the experimental data and the modelisation used to establish the dispersion equation  of the planar waveguide. Thus we conclude that this elaboration method is not adapted to realise a step-index waveguide. The films having undergone a slow cooling inside the oven (figure\ref{fig6}b) present a minimum for $m=0$, which corresponds to a refractive index close to $2.22$ and a thickness of approximatively $0.87\mu m$. In the case of this heat treatment, the films realised correspond well to the step-index waveguide model.

The homogeneity of the films and the reproducibility of the elaboration method have been studied
with M-lines spectroscopy, too. Two series ($S_1$ and $S_2$) of six films, elaborated under the same conditions(composition of the solution, parameters of spin-coating, temperature and duration of the heat treatment,\dots), have been investigated. In the case of these films the synchronous angles could be determined with a precision of only 0.1$^\circ$ due to rather flat peaks in the M-lines spectrum. Using the error calculation described in the previous section, we find an uncertainty of $\pm3.10^{-3}$ for the refractive index and $\pm8.10^{-3}~\mu m$ for the film thickness.

\begin{table}[htbp]
\caption{Mean values and dispersion of the refractive index and the film thickness of the two series $S_1$ and $S_2$.}
\centerline{
\begin{tabular}[t]{|c||*{2}{c|}}
\hline
Series&$<n>$&$max\mid n-<n>\mid$\\
\hline
$S_1$&$2.224$&$6.10^{-3}$\\
\hline
$S_2$&$2.221$&$7.10^{-3}$\\
\hline
Series& $<d>(\mu m)$&$max\mid d-<d>\mid$\\
\hline
$S_1$&$0.92$&$1.10^{-2}$\\
\hline
$S_2$&$0.87$&$3.10^{-2}$\\
\hline
\end{tabular}
}
\label{table1}
\end{table}
In order to determine the lower limit of the measurement accuracy of our experimental set-up, the films were measured many times at the same point. The maximum difference of refractive index and thickness for all measurement are respectively $2.2.10^{-3}$ and $5.10^{-3} \mu m$. These values are consistent with the uncertainties previously calculated. The mean values and the dispersion of the refractive index and the thickness of the two series of samples are presented in Table \ref{table1}. One can see, that the values are almost identically for both series, and the differences are smaller than the dispersion within one serie. From this overlaping of results between the two series, we deduce a reproducibility of the elaboration process, of the order of $5.10^{-3}$ for the refractive index and $20~nm$ for the thickness of the films.

\section{Conclusions}
In the present paper, a consistent method for indexing the peaks of M-lines spectrum obtained from PZT thin films was proposed. We have shown that the classical indexing criterion sometimes fails, and we proposed a method to test the consistency of the spectra. This method provides good results and can be applied to films which are not too far from the ideal step-index waveguide. By successively measuring  identical points of a sample, the sensibility of the M-lines set-up could be determined. The reproducibility of the elaboration process was quantified. In a preliminary study of the optical properties of spin-coated PZT thin films we have shown the crucial influence of the heat treatment procedure. The proposed method will also allow the study of other elaboration parameters.

%
%
%
%
%


\end{document}